\documentclass[12pt,letterpaper]{article}
\usepackage{amsmath,pgf,pgfarrows,pgfnodes,float,appendix, hyperref,scalerel,amssymb}
\usepackage{graphicx}
\usepackage{subfigure}

\usepackage[margin=0.9in]{geometry}
\usepackage{pgfplots}
\usepackage{amsmath}
\usepackage{tikz}

\title{{\rm\footnotesize \qquad \qquad \qquad \qquad \qquad \ \qquad \qquad \qquad \ \ \ \ \ \                   }\vskip.5in  Entropy and Black Holes in the Very Early Universe}
\author{Tom Banks\\ 
Department of Physics and NHETC\\
Rutgers University, Piscataway, NJ 08854\\
E-mail: \href{mailto:tibanks@ucsc.edu}{tibanks@ucsc.edu}
\\
\\
Willy Fischler\\
Department of Physics and Texas Cosmology Center\\
University of Texas, Austin, TX 78712\\
E-mail: \href{mailto:fischler@physics.utexas.edu}{fischler@physics.utexas.edu}}
\date{}
\begin{document}
\maketitle

\begin{abstract} 
Model independent arguments following from the Covariant Entropy Principle imply that causal diamonds in the very early universe were entirely filled with a single equilibrated system with finite entropy.  A universe where this condition persists forever has no localized excitations. Our own universe appears to be headed toward such a state.  Within a few hundred times its current age it will approach a state where our local group of galaxies sit in empty de Sitter space. Eventually, the local group collapse into a black hole, which evaporates.  Localized excitations in de Sitter space are low entropy constrained states of the vacuum ensemble.  The origin of these constraints must be in the early universe: the apparent horizon must expand after some initial period, in a constrained state that is the origin of all localized excitations in the universe.  We argue that in global FRW coordinates, this corresponds to slow roll inflation that ends in a dilute gas of tiny black holes, with mass determined by the inflationary scale.  We then review arguments that these black holes can account for the Hot Big Bang, baryogenesis, a distinctive pattern of CMB fluctuations, and possibly primordial black hole dark matter consisting of larger black holes that survive until the matter dominated era. The more complicated question of whether these small black holes can evolve in a way that is consistent with all observational constraints requires computer simulations that have not yet been done.
\end{abstract}

\section{Introduction}

A causal diamond in a Lorentzian space-time is the intersection of the backward lightcone of a point and the forward lightcone of a point in its causal past. The geodesic between the points is the timelike trajectory of maximal proper time, and the holographic screen of the diamond is the maximal area $d - 2$ surface in the null foliation of the diamond's boundary.  The {\it covariant entropy principle} (CEP) is a generalization\cite{fsb} of the Bekenstein-Hawking-Gibbons formulas\cite{bhg} for the entropies of black holes and of de Sitter(dS) space to arbitrary diamonds in arbitrary space-times.  It was anticipated in the work of 't Hooft\cite{tH} and in Jacobson's 1995 derivation\cite{ted95} of Einstein's field equations from the first law of thermodynamics.

In quantum field theory (QFT) each diamond is endowed with an operator algebra whose Heisenberg evolution between time slices that remain inside the diamond closes on itself.  The entanglement entropy of any state on this algebra with the system\footnote{This statement is true for any state on the full algebra for which the integrated stress tensor does not scale with the maximal volume of spacelike slices in the diamond.} in any larger diamond, is infinite, but proportional to the area of the holographic screen.  It is widely assumed that in a theory of quantum gravity, these infinities will be regularized by the finite Planck length.  The CEP simply implements this suggestion and uses the black hole examples to determine the normalization\cite{ted2015} of the area/entropy relation in terms of the Planck area. The CEP has been "derived"\cite{BDF} by Euclidean path integral methods similar to those\cite{GH} used to compute black hole entropy.

The purpose of this essay is to show that this simple principle, combined with the principles of statistical mechanics, has profound implications for the very earliest history of the universe, at least if we impose the condition that the final state of our universe will be empty dS space\footnote{There have been lots of claims in the literature that quantum gravity does not admit a stable dS space.  Only those claims which claim the instability sets in on time scales shorter than the time for our local group of galaxies to collapse into a black hole would modify the predictions of this paper for times before that collapse occurs .}.  As we'll review, this implies that any state of the universe, which contains localized excitations, is a low entropy constrained state. In a universe with no localized excitations, the entropy of the state in any causal diamond is always maximal and there is no interesting cosmological evolution.  If, as appears to be true in our own universe, the asymptotic future state has finite entropy then the results of\cite{fs} show that the system is uniquely modeled by a flat FRW universe with scale factor
\begin{equation} a(t) = \sinh^{1/3} (3t/R) , \end{equation} where $R$ is the asymptotic dS radius.

The absence of localized excitations means that this is not a good model of our own universe, but it holds lessons for us nonetheless.  In particular\cite{holocosmo}, if we want a quantum theory to actually describe the flat $p = \rho$ FRW cosmology, the time dependent expectation value of the Hamiltonian should satisfy the Friedmann equation
\begin{equation} \langle t | H(t) | t  \rangle = K V(t) t^{-2} , \end{equation} where $V(t)$ is the volume of the maximal spatial slice of the causal diamond along an FRW geodesic at cosmological time $t$.  Note that we've allowed time dependence both in the state and in the Hamiltonian, as always in the Schrodinger picture for a time dependent Hamiltonian.
The $p = \rho$ cosmology saturates the Covariant Entropy Bound, so we should imagine the time averaged state is an equilibrium state of $H(t)$, a density matrix concentrated in the part of the spectrum where the density of states is maximal.

We should also impose the equation of state of $p = \rho$, namely that $\rho \sim \sqrt{\sigma} $, the entropy density.  A solution of these constraints is that, for large enough $t$, $H(t)$ approaches the Hamiltonian of a $1 + 1$ dimensional CFT, with central charge proportional to the apparent horizon entropy $t^2$.  The CFT has a low UV cutoff $M \sim 1/t$ and lives on an interval of length $L$ with $ML$ large and independent of $t$.  Thus we have
\begin{equation} t^{-2} \sim c M^2 L t^{-3} \sim c^2 (ML)^2 t^{-6} . \end{equation} 

When we first made this observation we were unaware of the seminal work of Carlip\cite{carlipetal}, which had proposed that the horizons of black holes were described by such a CFT.  That work was motivated by the invariance of the near horizon geometry under a Virasoro subalgebra of the algebra of infinitesimal diffeomorphisms.  Using the symplectic structure of the Einstein-Hilbert Lagrangian, the Poisson algebra of these generators has a central charge and a classical value of the $L_0$ generator, which reproduce the Bekenstein-Hawking entropy for all Minkowski black holes, if one uses the Cardy formula.  In fact, this derivation works for the cosmological horizon of dS space, or any causal diamond smaller than the radius of curvature in maximally symmetric space-times.  In\cite{tbkz} we conjectured that this cutoff $1 + 1$ CFT is a general signature of the entanglement spectra of causal diamonds\footnote{The analysis of \cite{carlipetal} is ultralocal on the co-dimension two holographic screen of a diamond.
For large causal diamonds or black holes in AdS space, the spectrum of quasi-normal modes includes sound modes, which indicate that there are more global contributions to the entropy.  The black hole entropy formula for such diamonds, validated by the AdS/CFT correspondence, shows that the entropy is actually dominated by the entropy of generalized sound modes.  In tensor network language, the horizon looks like a lattice of $1 + 1$ dimensional CFTs of large central charge, with short range couplings between them. }.
This picture suggests that the Hamiltonian describing the equilibrium states of dS space is a cutoff $1 + 1$ dimensional CFT, identical to that describing a Minkowski black hole of the same radius. The Hamiltonian should also incorporate the property of {\it fast scrambling}\cite{hpss} of information on the dS horizon.  This is achieved if the geometry of the holographic screen is related to the target space of the $1 + 1$ CFT and the dynamics is invariant under area preserving diffeomorphisms of the screen.  The finite total entropy then implies that the algebra of measurable functions on the screen must be finite dimensional.

We'll argue that the maximal entropy initial condition with localized excitations is an inflationary universe, which evolves after a period of slow roll into a dilute gas of black holes with average mass given by the inflationary horizon size.  We call these {\it inflationary black holes}, IBHs. Since these are finite entropy systems, there will be fluctuations in both their mass and angular momentum, which will show up as scalar and tensor fluctuations in the CMB.   CMB data indicates that the black hole mass is of order $10^6$ in Planck units.  These black holes therefore decay rapidly and their decay leads to the hot Big Bang.  The decay also leads to a second source of primordial gravitational waves, in addition to those coming from intrinsic angular momentum fluctuations.
In addition the decay leads to violation of CPT proportional to the derivative of black hole mass.  Given order 1 CP violation, this can generate the baryon to entropy density ratio required to fit the data.   

The period between inflation and the HBB is a matter dominated universe so fluctuations grow. It is plausible that some of the inflationary black holes (IBHs)  have merged by the time the HBB occurs, and will decay long after the HBB.  It is simple to estimate the number of black hole mergers that will be required in order to produce black holes that survive until they dominate the HBB radiation density, and in addition that this second matter dominated era will begin at a temperature of $1$ eV.  It turns out that we only require a probability of $\sim 10^{-24}$ for mergers of order $10^{6}$ IBHs in order to achieve the correct cross over density.  These models may produce a scenario in which the dark matter consists of primordial black holes (PBHs) at the time of matter radiation equality.

Thus, a simple, almost parameter free, model can reproduce all of the observational signatures of early universe cosmology, CMB fluctuations dominated by the scalar modes, the HBB, the baryon to entropy ratio, and possibly the existence of PBH dark matter .   Of course, just like conventional inflation models, the detailed shape of the CMB spectrum depends on details of the slow roll end of inflation.  In models based on the CEP, there are some mild constraints on this coming from the requirement that the universe expands rapidly enough during slow roll to prevent the assumed constraints from being erased, reverting to the featureless maximal entropy cosmology, but there does not seem to be any barrier to fitting the data.  The deviation of the spectrum from scale invariance comes entirely from a prefactor $\epsilon^{-2} (t(k))$ in the scalar power spectrum.  $\epsilon$ is the slow roll factor.

The problem that remains to be solved is growth and merger of the PBHs during the second matter dominated era.  This is a complex nonlinear problem which requires computer simulation.  When compared with conventional galaxy formation simulations one must add the fact that the PBHs have finite size and that their merger forms larger black holes.   Such a simulation is the only way to tell whether the model produces a spectrum of PBH dark matter that is compatible with all observational constraints, and whether these PBHs could be all of the dark matter.  The simulation might also shed light on the mechanism for the growth of the central black hole cores of galaxies and make predictions for the spectrum of black holes formed from primordial mergers rather than collapse of stars.

\section{The Beginning of the Universe}

Consider a causal diamond at some proper time $t$ from the beginning of the universe.  The area of the diamond is finite.  We will also insist that the space-time geometry is such that there is a regime of $t >> 1$, where the radius of curvature of the geometry is much larger than $t$\footnote{In the Holographic Space-time models of\cite{holocosmo} this requirement is built into the choice of the time evolution operator, and does not constrain the initial state at all.  Since we don't want to tie our considerations to particular models, we have to make this an explicit assumption in this essay.}.
Thus, the CEP implies that the system has a maximal entropy $A/4\sim t^2 $. Here and henceforth we let $G_N = L_P^{-2} = 1$ .The scaling relation between area and proper time is appropriate if the diamond is small compared to the "radius of space-time curvature".  In quantum mechanics, this implies that the Hilbert space is finite dimensional.  Saturating the entropy bound means that the initial state is chosen randomly, rather than from some low dimension subspace.  The only natural time scale, other than the Planck scale, is $t$ itself, so the typical time scale we can find within the diamond is of order $t$\footnote{Of course, the extreme degeneracy of eigenvalues implied by a number of states of order $e^{t^2} $ with a Hamiltonian of order $1/t$ would imply much longer "recurrence" time scales, but the actual dynamics of the universe on time scales much longer than $t$ involves a different Hamiltonian, acting on a much larger Hilbert space.  Recurrences occur in cosmology only long after the final, empty dS state, is reached.  They can be avoided in a theoretical model, with no consequences for any conceivable observational check, by modifying the time dependent Hamiltonian at times much larger than the dS radius, and much smaller than the recurrence time.}.  Note that in QFT, the operator algebra of a diamond evolves to itself under Heisenberg time evolution, but that evolution is not generated by a unitary operation in the diamond algebra.  The CEP guarantess that there is such a unitary, $U(t)$, but that operator will not be the same one that generates evolution on a larger diamond.   The Hamiltonian $i {\rm ln}\ U^{-1} (t) U(t+ 1)$ will be time dependent.

Proper time is time as measured along the geodesic in the diamond.  If we think about hypothetical "local observers traveling on the geodesic" in this (tiny) diamond, their natural coordinate system is the {\it cosmological Milne} coordinate system 
\begin{equation} ds^2 = a^2 (s)  \frac{t^2}{4} (- ds^2 + (s - t)^2 [dx^2  + \sinh^2 x d\Omega^2])\ \ \ \ 0\leq s \leq t. \end{equation}  $s = 0$ is the place where $a(s)$ vanishes. Systems near the boundary of the diamond evolve very slowly w.r.t. to proper time on the geodesic (the Milne red-shift), while those localized near the geodesic have "typical" $o(1)$ time evolution.  Einstein's equations tell us that a local concentration of energy of order $E$ will form a black hole of radius $\sim E$ centered on the geodesic, if $E > 1$.  

Now we want to introduce a fundamental principle, which is most simply approached by studying the Schwarzschild black hole metric (SdS) in de Sitter (dS) space
\begin{equation} ds^2 = - f(r) dt^2 + \frac{dr^2}{f(r)} + r^2 d\Omega^2 , \end{equation}
\begin{equation} r f(r) = - R^{-2} (r - R_+) (r - R_-)(r + R_+ + R_-) , \end{equation} where
\begin{equation} R_+ R_- (R_+ + R_-) = \frac{2M}{R^2} , \end{equation}
\begin{equation} (R_+ + R_-)^2 - R_+ R_- = R^2 = R_+^2 + R_-^2 + R_+ R_- . \end{equation}
The sum of the entropies of the two horizons, according to the CEP, is
$\pi (R_+^2 + R_-^2) = \pi (R^2 - R_+ R_-) $, which means that the black hole has an entropy deficit $\pi R_+ R_-$ relative to empty dS space.  This means 
\begin{itemize} 
\item Empty dS space is a high entropy mixed state, rather than a pure state.
\item The probability of finding a black hole with radius small compared to the dS radius, sitting at the origin of a static patch, is approximately $e^{ - 2\pi M R}$.  This is a derivation of the temperature of empty dS space from the CEP and classical physics, with no necessity for a quantum field theory calculation.  The fact that the Euclidean time of empty dS space has this periodicity is independent confirmation of this fact.
\item A localized object of mass $M$, which is not a black hole, has an even smaller probability, because it does not have the $\pi (2M)^2 $ contribution to its entropy.
\item It follows that most of the states are localized near the horizon, which is consistent with the fact that in cosmological Milne coordinates they have highly redshifted energies.  The Boltzmann formula tells us that a typical state in the vacuum ensemble has energy of order $1/R$, measured along the geodesic. 
\end{itemize}

The "prediction" of the CEP that dS space has a finite temperature, was spectacularly, if anachronistically verified by the famous calculation of Gibbons and Hawking\cite{gh1} which shows that a static object in dS space is constantly bombarded by thermal massless particles of momentum $\sim (2\pi R)^{-1}$, and therefore cannot remain static.  It will perform a random walk away from the origin by a distance $d$ in a time of order $mR d^2$ in Planck units and eventually will be swept out of a given mathematical causal patch by the Hubble flow.  This is the mechanism by which the system returns to equilibrium, starting from the constrained state with a localized object.  Note that the time for equilibration of a system initially at rest on the geodesic scales like $R^2$ in this naive calculation.  Systems moving relative to the geodesic equilibrate in a time of order $R$.

If we consider a very massive object like our local group of galaxies, we might want to neglect this deviation from geodesic motion, but that object will radiate massless particles out to the horizon and eventually collapse into a black hole.  The collapse time is independent of $R$ as long as the black hole size is much smaller than $R$.  After that time, no theoretical model has any observational relevance.   If we've accepted the necessity of describing local physics by time dependent Hamiltonians, no observation can distinguish a model of eternal dS space from one in which the dS era ends sometime after the collapse of the local group.  All paradoxes related to an assumed eternal persistence of dS space are irrelevant to any conceivable observation.

Now let us apply these insights to cosmology using two principles.  We would like to maximize the probability of the choice of initial state, and we also want to take into account the fact that the universe we inhabit appears to be approaching a dS space.
We can use the above observations in several ways.  First of all, we conclude that a universe with any localized objects in it at all, is a fine tuned initial condition.  Secondly, we conclude that the most probable localized objects to be found are black holes.   

We've already observed that a cosmology that saturates the CEP at all times has scale factor $\sinh^{1/3} (3tH_I)$ and contains no localized excitations at any time.   Let us assume that after of some number of e-folds, this is modified by a slow expansion of the horizon, but in a constrained state corresponding to the appearance of localized excitations.  Consider this from the point of view of a given geodesic G1.  The causal diamond along that geodesic has a fixed area for some number of e-folds and then begins to expand.  Consider the event at which some other geodesic, G2, crosses the past boundary of that extended diamond, as shown in Figure 1.   
\begin{figure}[btp]
\begin{center}
\includegraphics[scale=0.45]{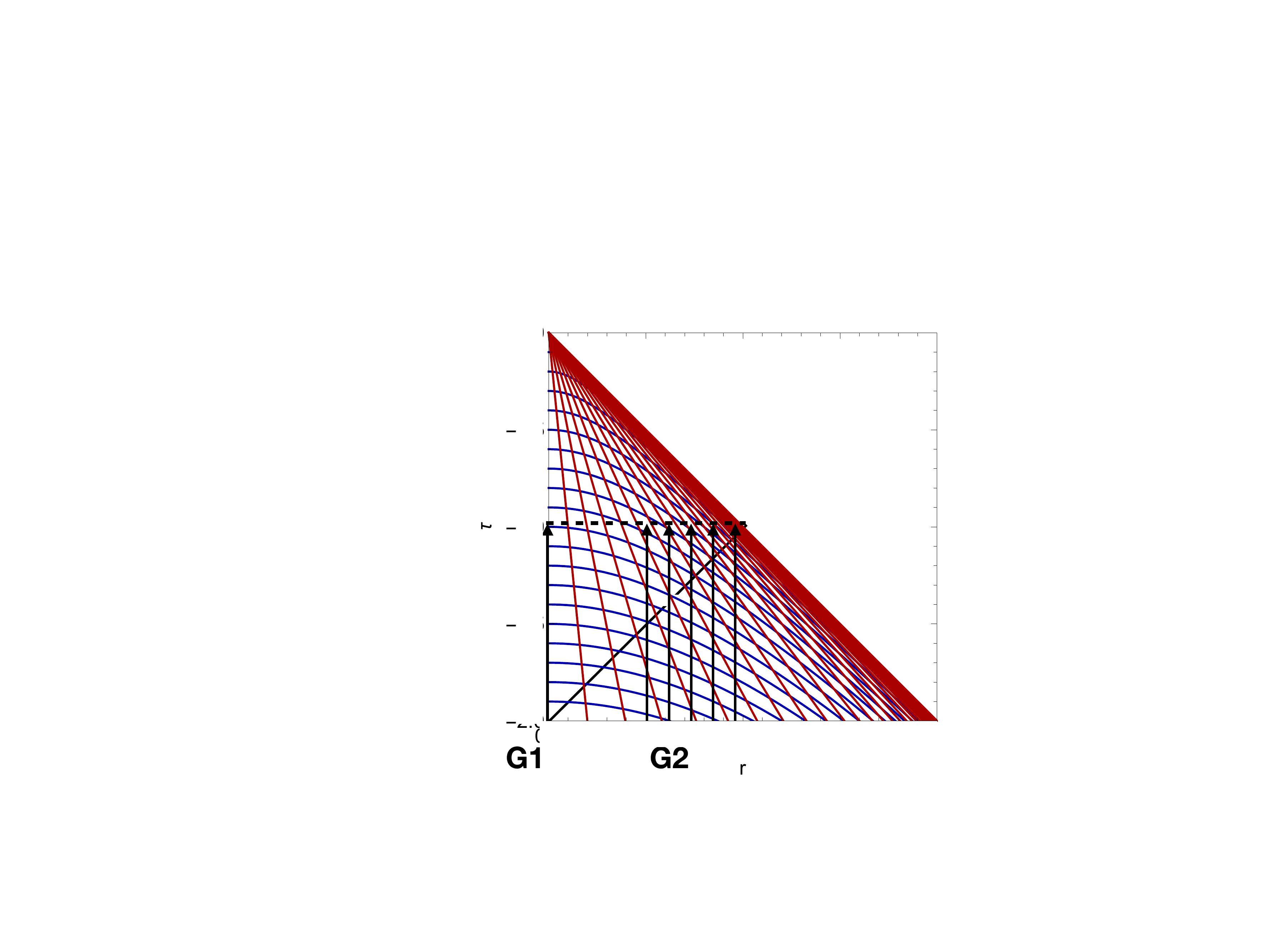}
\label{fig:hstuniv}
\vspace{-0.5cm}
\caption{A Geodesic 2 enters the past horizon of Geodesic 1 at an FRW time when the universe is still inflating.  It is an isolated, finite entropy system, in thermal equilibrium, and so appears to be a black hole.  The dashed line marks the end of the FRW slow roll period, when the universe becomes dominated by a dilute gas of black holes. The blue lines are equal time surfaces in cosmological Milne coordinates.}
\end{center}
\end{figure}

In FRW coordinates, that event is in the past, while in backward Milne coordinates on the flat space to which the FRW geometry is conformal, it is contemporaneous\footnote{FRW and Milne time agree along the geodesic in a particular diamond, but they slice space-time differently.}.  In general, for far away events, the geodesic G1, at the FRW time contemporaneous with the horizon crossing event, was still undergoing inflation.  Therefore, the "localized object" seen at the event by G1 could be nothing but the inflationary horizon of G2\footnote{In HST models, this choice of equal treatment of all geodesics, or what we generally call homogeneity, is part of the mathematical construction of the model and does not require fine tuning of initial conditions.}.   This localized object is a high entropy equilibrated quantum system with a certain geometrical size.  It will have a finite temperature.   To a detector at G1 it will appear to be a black hole.   

There are two important points here.  First, we're assuming that the horizon of empty dS space has dynamics identical to that of a black hole with the same area.  This is a natural consequence of the description of horizons advocated in\cite{carlipetal}. Secondly, we're describing the universe simultaneously by Hamiltonians adapted to two different time-like geodesics.  The identification of the inflationary horizon from the point of view of G2 as a black hole horizon from the point of view of G1, is a coarse grained version of the requirement that the density matrices assigned to the maximal causal diamond in the overlap of the diamonds of two different trajectories, have the same entanglement spectra.  

The burden of these remarks is that, from the point of view of the CEP, a slow roll inflationary model naturally ends up as a dilute gas of black holes whose Schwarzschild radius coincides with the inflationary horizon size, as long as the universe expands fast enough for the black holes to avoid merging into a single horizon filling hole. We'll call these Inflationary Black Holes, IBHs.  It's important to emphasize that the inflationary scalar field is, in this view, an artificial construct.
One starts with a scale factor $a(t) =\sinh^{1/3} (3 H(t) t) $ with $\epsilon \equiv \frac{\dot{H}}{H^2} $ small and {\it defines} $p$ and $\rho$ from the Friedmann equation.  The underlying quantum model is not a quantum field theory but a system with a time dependent\footnote{As discussed above, time dependence is forced on us by our choice of cosmological Milne coordinates, as well as by the implementation of causality without quantum field theory.} Hamiltonian in a finite dimensional Hilbert space.  The growth of the size of the diamond Hilbert space with proper time, determines the slow roll geometry. However, if $p + \rho \geq 0$ then we can find a scalar field Lagrangian whose classical dynamics maps out the same geometry.  The null energy condition is just the statement\cite{ted95} that entropy is always increasing.  In the underlying quantum dynamics this occurs simply because more degrees of freedom are coupled to the system as DU time increases.

We get a fundamental constraint on the slow roll geometry by insisting that the horizon increases rapidly enough to prevent the constrained state from coming back to equilibrium.  The equilibration time of a system with Hubble constant $H(t)$ is $H^{-1} {\rm ln}\ H^{-1} $ and this must be longer than the time it takes the horizon to expand by a factor of order $e$.  Thus
\begin{equation} \epsilon > \frac{c}{{\rm ln}\ H^{-1}}.\end{equation} Here $c$ is a constant, nominally of order 1, that characterizes the particular time dependent Hamiltonian describing scrambling in the slowly expanding diamonds. Since the typical energy scales in the scrambling Hamiltonian are of order $H$, the scrambling bound of \cite{mss} tells us that $c$ cannot be too large.   Unitarity puts
no constraints on the form of the fast scrambling Hamiltonian, or on the rate to which we add degrees of freedom to it\footnote{We caution our readers that these added degrees of freedom are merely being transferred from another subsystem describing evolution of the universe outside the causal diamond.  There is no violation of unitarity involved.}.   Relativity merely tells us that if we look at cosmology from the point of view of a different timelike trajectory then the density matrices predicted by the two models for the common subsystem corresponding to the overlap between diamonds must have the same entanglement spectra.  We have used this principle to conclude that the intersection of another FRW trajectory with the past boundary of "our" causal diamond is interpreted as a black hole entering our horizon.  Relativity has no constraints on the slow roll era, but becomes important once we have localized objects in the universe. In figure 1, this occurs at the FRW conformal slice halfway between the origin and the position $\eta_0$ of the pole in the scale factor describing the dS era.    

Thus, apart from the condition $p + \rho \geq 0$ for the fluid stress tensor derived from our choice of $a(t)$ and the scrambling constraint on $\epsilon$, there seem to be no theoretical constraints on our choice of slow roll metric.   Constraints coming from considerations of low energy scalar field Lagrangians are not applicable to models based on the CEP, because the inflaton field is a phenomenological construct unrelated to scalar fields describing low energy scattering amplitudes in Minkowski string theory.   Note that CMB data suggest that $\epsilon \sim 0.1$ and $H \sim 10^{-6}$, so the scrambling bound is close to being saturated if $c \sim 1$.  This is consistent with the principle that the most probable initial conditions for the universe are those in which the density of localized black hole excitations at the end of inflation is just below the threshold at which they would merge to form the maximal entropy universe with a Hubble constant only slightly smaller than $H_I$.  

From a phenomenological point of view, the constraint of no return to maximal entropy evolution is an upper bound on the density of black holes at the end of inflation.  The black holes are finite quantum systems with equilibration times of order the Schwarzschild radius $m {\rm ln}\ m$.  As such, they undergo entropy fluctuations.  Conventional statistical mechanics suggests that these are of order $\sqrt{S}$ and we've recently shown\cite{tbkz} that the precise coefficient (assuming Einstein gravity is a good approximation) is
\begin{equation} (\delta K)^2 =  S. \end{equation} Here $K$ is the modular Hamiltonian, whose expectation value is the black hole entropy.  This leads to a precise formula for black hole mass fluctuations.  This is translated into fluctuations of the gauge invariant metric parameter 
$\zeta$ by the standard formulae
\begin{equation} \zeta = \delta \tau = \frac{\delta H}{\dot H} = \frac{\delta H}{\epsilon H}.  = \frac{H_I^2}{\pi \epsilon }\end{equation}
Here $\delta \tau$ is the proper time fluctuation between two time slices in co-moving gauge.

Note the prediction for the power spectrum is larger by a factor of $\epsilon^{-1}$ than the predictions for scalar fluctuations obtained by quantizing Einstein's Lagrangian\cite{maldaetal}.  The reason for this is that quantum fluctuations of the scale factor in comoving gauge, treated as a quantum field, are suppressed by a factor of $\sqrt{\epsilon}$ relative to the
predictions of generic statistical mechanics, while black hole fluctuations $\frac{\delta m}{m}$ are independent of $\epsilon$.  Note also that the factor of $H$ in the fluctuations is $H_I$ rather than the slow roll Hubble factor $H(t)$.  Again this is because black holes are isolated quantum systems, whose fluctuations do not depend on the scale factor of the geometry in which they're embedded.  

Tensor fluctuations can be viewed as fluctuations in which a black hole gains or loses angular momentum by emitting gravitational waves.  The rotating black hole entropy formula tells us that for fixed mass the angular momentum distribution is centered around zero with a Gaussian width $ \sqrt{2\pi/M^2}$.  Thus, the tensor fluctuations are down by a factor of $\epsilon^2 k_L$ relative to the scalar power.  One should be able to calculate the order one constant $k_L$ by comparing the angular momentum fluctuations in a scale invariant distribution of primordial gravitons with those predicted by the rotating black hole entropy formula.  We have not yet performed this calculation.

The approximate $SO(1,4)$ invariance of the scalar and tensor curvature two point functions can be understood as follows.  At each angle, the black hole fluctuations are described by an approximately scale invariant Hamiltonian, as described above.  Thus, all we need do is build in approximate rotation invariance, to get the predictions of $SO(1,4)$ invariance.  The local generators of $SL(2,R)$ at fixed angle are given by
$L_0 (\Omega)$ and $\vec{\Omega}\cdot \vec{L}_{\pm} (\Omega)$ and the $SO(1,4)$ generators are the integrals of these operators over $\Omega$.

Deviations from that scale invariant behavior are down by powers of $H_I$ in Planck units apart from the factors of $\epsilon$ in front of the scalar power.  Thus, the tensor spectrum is predicted to be flat and the tensor to scalar ratio quite small, of order $\epsilon^2$.  However, we'll see in a moment that models based on the CEP have another contribution to the primordial gravitational wave spectrum coming from decay of the black holes.  This will have a spatial distribution mirroring that of the scalar power spectrum and an amplitude that is the scalar power multiplied by the inverse of the number of particle species with mass below the decay temperature.  That temperature is of order $10^{8}$ GeV, so that suppression could easily be a factor of $10^{-3}$.    

These arguments do not specify the size of the inflationary horizon/average black hole mass.   It is easy to argue that it must be much larger than Planck scale.  The immediate post inflationary universe is matter dominated, so any fluctuations will grow and become non-linear.  Since our model of the mass distribution is a collection of black holes, the nonlinear dynamics implies black hole collision, merger and growth.  If this process is too efficient, then we will end up with a system that comes back to equilibrium.  We will have a universe which is either maximal entropy, or consists of huge metastable black holes for its entire history.  The only parameter we have to control the efficiency of the merger process is the amplitude of primordial inhomogeneity, so this must be small, which implies that the inflationary horizon was much larger than Planck scale.  Of course, we have assumed this, along with the CEP, in giving a statistical description of the individual black holes. Comparing our model to CMB data show us that
\begin{equation} m\epsilon = 10^5  , \end{equation} so that $m \sim 10^6$ for typical CMB fits to $\epsilon$.

During the dilute black hole gas (DBHG) era of the cosmos, fluctuations grow like
\begin{equation} \frac{\delta\rho}{\rho} = (m\epsilon)^{-1} (\frac{\rho_I}{\rho})^{1/3} . \end{equation} If the black hole number density just after the end of inflation is $\sim m^{-3}$ (so that the energy density $\rho_I$ is $m^{-2}$, consistent with conventional inflationary expectations), 
then the nonlinear regime begins when
\begin{equation} m \epsilon =  (\frac{1}{m^2\rho})^{1/3}, \end{equation}
\begin{equation} \rho = \epsilon^{-3} m^{- 5}  . \end{equation}
This corresponds to a time $t \sim m^{5/2} \epsilon^{3/2}$ after the end of inflation.  Note that it is much less than the black hole evaporation time $t_d \sim g^{-1} m^3$, where $g$ is the number of species with mass below the black hole temperature $10^{13}$ GeV, so long as $g < m^{1/2} \epsilon^{-3/2} \sim \frac{10^5}{r_I}$.  $r_I$ is the ratio of tensor to scalar fluctuation modes produced by inflation.  

If we start out with an IBH number density just below that at which the universe would evolve to a maximum entropy state, then this observation means that we are in a regime where some black holes will merge and become more stable, while the rest decay, giving rise to the Hot Big Bang.  We are again following the rule that we choose maximal entropy initial conditions compatible with the existence of localized objects in the universe.   The task of determining the spectrum of relic black hole masses that survive after most of the IBHs decay into radiation is a complicated one, which can only be accomplished by computer simulation.  We will see however that only a tiny fraction of the IBHs have to merge in order to form enough meta-stable black holes to dominate the energy density of the universe at the observed crossover between matter and radiation.  The necessary fraction, $10^{-28} $, is so small that we may be forced to confront one of the fundamental issues that arise in any quantum mechanical theory of cosmology.

Quantum mechanics makes only probabilistic statements and the verification of a probabilistic theory requires one to "rerun the experiment a large number of times" in order to use the frequentist interpretation of probability.  In cosmology we get to run the experiment only once.  A model that predicts coarse grained observations correctly with probability close to one is the cure for this conundrum. In the kind of model we are considering here we might not have that luxury.  Suppose we were to find that the model did not have a particle physics candidate for dark matter, and that the probability of forming black holes that could persist down to temperatures of order $1$ eV before the Big Bang was $.00000001$.  One might be tempted to say that the model did not agree with observation.  However, once we know that we only need a probability of  $10^{-28} $ to have mergers of $\sim 10^5$ IBHs into such meta-stable PBHs in order to fit the data, we are inclined to accept that the character of our universe depends on a "quantum mechanical accident".  As a practical matter, since it is unrealistic to expect computer simulations accurate to the $10^{-8}$ level, we probably have to accept that we will be unable to prove conclusively that our model predicts the observed universe with high probability.

Before proceeding to a more detailed analysis of Primordial Black Hole (PBH) Dark Matter in models based on the CEP, we want to discuss baryogenesis.  The reheat temperature in these models is on the order of $10^{8} - 10^9$ GeV so many particle physics mechanisms for baryogenesis don't work.  However the decay of the IBHs may have all of the ingredients necessary for baryogenesis\cite{tbwfbaryogen}.  The decay process does not preserve baryon number, and the variation of the black hole mass during decay violates CPT.  Thus, the only missing ingredient is CP violation.  We know that CP is violated in our universe, and extant data suggests that the only reason it does not appear of order $1$ is related to the masses and mixing angles of the third generation of quarks.  Assuming that black hole decay matrix elements have order $1$ CP violation, the total baryon number produced in the decay of a single black hole should be\cite{tbwfbaryogen} 
\begin{equation} \frac{dB}{dt} \sim - \epsilon_{CP} m^2 \frac{dT_{BH}}{dt} \sim - \epsilon_{CP} \frac{d(m/M_P)}{dt} . \end{equation}  The amount of baryon number produced in a single black hole decay is thus
\begin{equation} \Delta B = - \epsilon_{CP} (m/M_P) , \end{equation} which has a universal sign.  The black hole gas is quite dilute at the time of reheating so the chance of baryons emitted in the decay of one black hole being absorbed by another is small.  Thus, the total baryon density produced in the decay is (in Planck units)
\begin{equation} \Delta b = - \epsilon_{CP} n_{BH} (m) . \end{equation} Here $n_{BH}$ is the number density of IBHs at the time of black hole decay.  Here we're assuming that the number density of those which coalesce and do not decay at a time of order $m^3$ after the end of inflation, is small.  At this time, the energy density is
\begin{equation} \rho_{RH} = m n_{BH} = g T^4 , \end{equation} where $g$ is the number of single particle massless states (with appropriate Bose/Fermi weighting) with mass below $\frac{1}{8\pi m}$.   The entropy density is
\begin{equation} g^{1/4} (m n_{BH}^{3/4} , \end{equation}
so the ratio of baryon density to entropy density is
\begin{equation} r_B = \epsilon_{CP} g^{-1/4} (m n_{BH}^{1/4}) . \end{equation}

At the end of the inflationary era, the IBH number density is bounded by 
\begin{equation} n_0 \leq \frac{3}{32\pi} m^{-3} , \end{equation} because a higher density of black holes would lead to immediate merger and the formation of a maximal entropy $p = \rho$ state.  The correct initial density is one which maximizes the probability of the assumed initial state given the constraint that localized objects persist until a time of order the dS Hubble time of our universe.  $n_{BH}$ is
\begin{equation} n_{BH} = n_0 \tau^{-2} , \end{equation} where $\tau$ is the black hole lifetime
\begin{equation} \tau =  \tilde{g}^{-1} m^{3} . \end{equation} Here $\tilde{g} = \frac{g}{512\pi}$
Thus
\begin{equation} r_B =  \epsilon_{CP} m^{-2} \tilde{g}^{7/4} n_0^{1/4} \leq  (3/32)^{-1/4} m^{-2} \tilde{g}^{7/4} =  (1.8)^{-1} 10^{-12} (1.8) \times 10^{4} \sim \epsilon_{CP} \times 10^{-8} . \end{equation}  We've used the value $m = 10^6$ given by the fit to the CMB, with the slow roll parameter $\sim 0.1$ and taken $g = 10^{4} $ Given the crudeness of our approximations, this is remarkably close to the required phenomenological value.

Next we note that black hole decay will produce a spectrum of gravitons whose spatial profile will mirror the scalar fluctuation spectrum.  Its power spectrum is down by a factor of $1/g$ relative to the scalars.  Depending on the actual value of $g$ and of the slow roll parameter, this could be the dominant contribution to primordial tensor curvature fluctuations.   The prediction of a two component contribution to tensor fluctuations, with characteristic spatial distributions, is a robust and unique prediction of models based on the CEP, but there is no reason to believe that the slow roll parameter and $g$ are in the fine tuned range where both components are easily observable.  Theoretical expectations for their values are such that neither component is in conflict with the current absence of a signal for tensor fluctuations.   

\section{Primordial Black Hole Dark Matter}

We've seen that density fluctuations in the IBH gas go non-linear at a time parametrically smaller than the black hole decay time, which signals the beginning of the radiation dominated era.  The task of calculating the spectrum and spatial distribution of black holes larger than the IBHs, which is produced by mergers is a difficult one.  Instead of attempting it, we will estimate how many black holes of mass $M$ are required in order to have a distribution of PBHs that will dominate the energy density of the universe at the observed crossover between radiation and matter dominated eras at radiation temperature $\sim 1$ eV $ = 10^{-28}$ in Planck units.

We will assume that the initial distribution of those PBHs is sufficiently uniform and dilute that PBH collisions are negligible.  The first problem to address is then the evolution of a black hole of initial mass $M$ in a radiation dominated universe.  This was first studied by Novikov and Zeldovitch\cite{nz} followed by work of Carter and Hawking\cite{ch}.  The conclusion of these studies is that if the initial radiation dominated energy density is much less than the Planck scale, then black holes do not typically grow very much during a radiation dominated era.  When the radiation temperature falls below the Hawking temperature, they decay at roughly the rate of decay in flat space.  The fixed line of solutions found in\cite{nz}, where the black hole grows to fill the horizon, is unstable.

We can use this simple scenario to understand what happens to an initial spectrum of black holes, as long as their masses are spaced widely enough.  The lightest black holes decay, leading to a slight\footnote{We're always assuming the total initial energy density in relic black holes is much less than the initial radiation energy density.}  rise in the radiation temperature.  The largest initial black holes are the only ones that have a chance to dominate the universe as the radiation energy density falls.  Thus, it makes sense to concentrate on scenarios with only a single species of relic black hole, with a mass large enough to last until the observed end of the radiation dominated era.

Let us review the chain of arguments behind the last sentence.  The largest relic black hole will increase its mass by absorption by an order one factor, because the initial conditions of our models are not finely tuned.  Smaller relic black holes will reheat the universe only slightly, and so will not significantly affect the lifetime of the largest relic. Finally, we want to choose the lightest possible maximal relic, because initial conditions with larger relics will be less probable.  As always, we are guided by the principle that we should choose the most probable initial conditions consistent with some coarse grained feature of the universe we observe.  In our present state of understanding, this means that the largest relic should not decay before the observed end of radiation domination, under the assumption that the relic black holes are the dark matter.  After that point in time, the problem of black hole growth becomes complicated by merger histories.  One could study the beginnings of the matter dominated era with linearized perturbation theory, but a complete understanding would require full blown computer simulations.  It is only then that one could assess the possibility that models based on the CEP predict PBH dark matter.

We will restrict ourselves to the simple task of finding the required post Big Bang number density of PBHs to give PBH dark matter at the very beginning of the matter dominated era.  The requirement is encapsulated in three equations

\begin{equation} 1 = \frac{\rho_{eq}}{\rho_{PBH} (\rho_eq)} =  \frac{\rho_{eq}}{\rho_{PBH} (\rho_{RH})( \rho_{eq}/ \rho_{RH})^{3/4}} . \end{equation}
\begin{equation} \tau_{PBH} \geq  ( \rho_{eq}/ \rho_{RH})^{- 1/2}= t_{eq}. \end{equation}
\begin{equation} \tau_{PBH}  =  \tilde{g}_<^{-1} X^3 m^3 . \end{equation}
The first is the statement that $\rho_{eq} $ is the energy density of PBH and radiation crossover.  The observed value, in Planck units, is $10^{-112}$.  We also recall that
\begin{equation} \rho_{PBH} (\rho_{RH}) = X r(X) \rho_{RH} . \end{equation} The PBH mass is $M = m X$ and $r(X)$ is the ratio of PBH to IBH number densities just before the Hot Big Bang. $t_{eq}$ is the cosmological time between the Big Bang and matter radiation equality.
The second equation is the statement that the PBH lifetime is at least as long as the time from the Big Bang to the crossover, and the third is the formula for the PBH lifetime.  $g_< < g$ is the number of particle species with mass $<< (8\pi M)^{-1}$.  

We obtain
\begin{equation} X r(X) = (\frac{\rho_{eq}}{\rho_{RH}})^{1/4} = 10^{-28} g^{-1/2} m^2 \sim 10^{-18} . \end{equation}
The bound on PBH lifetimes is 
\begin{equation} m^3 X^3 g_<^{-1} > 10^{56} g m^{-4} , \end{equation} or
\begin{equation} X > (g_< g 10^{14} )^{1/3} \sim 10^6 . \end{equation}  This implies
$r(X \sim 10^6) \sim 10^{-24} , $ with even smaller fractions for higher mass PBHs. 
As promised, the fraction of IBHs that must coalesce into PBHs in order to have PBH dark matter dominate the universe at the observer radiation/matter crossover is extremely small.  Although it takes of order $10^6$ IBH mergers to form a single $PBH$, the required ratio of densities is so small that such a merger history is not implausible.  It may prove very difficult to build computer simulations accurate enough to test this hypothesis.  Note that it's likely that the smallest allowed value of $X$ has the highest probability.

Since our models are quantum mechanical, they can at best calculate a probability distribution $P[r(X)]$ that PBHs of mass $M$, with $\rho_{eq}$ fitting the data, can be formed by the coalescence of IBHs.  In order to estimate whether these models predict a high probability for PBH matter domination at the observed crossover density, one must have an accurate calculation of $P[r(X)]$ for $X > 10^{6}$.  Note that the required fraction of PBHs falls like $X^{-1}$ for large $X$, but it's likely that $P[r(X)] \neq 0)$ falls more rapidly.  It might be possible to find an analytic asymptotic formula for $P[r(X)]$ for large $X$, but one would also need to understand the corrections to the asymptotic behavior near $X = 10^{6}$.

There is a more profound question raised by these considerations.  As long as the functional probability distribution for $r(X)$ allows any pair $X, r(X)$ that satisfies these conditions we can say that our model allows for the possibility that we have PBH dark matter. Our observations can only, {\it in principle} cover what happens in a single cosmological history, even in the context of a multiverse model. Our observations cannot rule out a theory that predicts our observed history with a tiny probability.  Multiverse models could be constructed, which allowed us to use this probability distribution to rule out CEP models, but there would be no reason to believe such a multiverse model unless it led to other verifiable predictions for our own universe, which were inextricably tied to its prediction that CEP models of PBH dark matter were incorrect.  We leave it to the reader to judge whether multiverse models with such verifiable predictions .

Models with $X \sim 10^{15}$ and a few orders of magnitude above this are consistent with all current constraints on PBH dark matter\cite{constraints}.  For those in the range $10^{6} < X <10^{15}$ we must solve the complicated problem of black hole merger and coalescence during a PBH matter dominated era, in order to determine whether signals from black hole decay can be used to rule these models out.  Furthermore, these models could be models of decaying dark matter, some of which have been invoked to resolve the Hubble tension and other problems of vanilla LCDM models\cite{ddm}.   The possibility of several species of black holes with masses spanning the range $10^{12} - 10^{21}$ in Planck units would lead to models of partially decaying dark matter, with considerable flexibility in fitting cosmological observations.

\section{Conclusions}

The models discussed in this paper are based on a couple of general physical principles and contain very few parameters.  The principles are the CEP and the related argument that localized objects are constrained states of a finite dimensional quantum system that lives on the instantaneous horizon of causal diamonds in the universe.  These lead to the general principle we have used throughout: choose cosmological initial conditions of highest probability in the space of quantum states, consistent with the existence of long lived localized objects.  We argued that these principles lead inevitably to a slow roll inflationary cosmology, which terminates in a dilute gas of black holes whose average Schwarzschild radius is the inflationary horizon.  The quantum statistical fluctuations in the individual black hole states lead to the observed CMB fluctuations.  Recent work\cite{tbkz} has enabled us to tie down the form of these fluctuations more precisely.   They fit the CMB data if we impose the additional requirement of approximate $SO(1,4)$ invariance.  This cannot be derived from our general principles, but the conjecture that black hole horizons are described by an approximate $1 + 1$ dimensional CFT, implies that fluctuations are scale invariant in each direction in the sky.  The predictions of these models are distinct from predictions of field theory inflation models and a measurement of tensor modes may be enough to decide which kind of model is correct. On a more theoretical level, field theory models violate the CKN bound\cite{CKN}\cite{BD} on the validity of field theory in finite causal diamonds, while models based on the CEP automatically satisfy that bound.  

Decay of the preponderance of the IBHs gives rise to the Hot Big Bang, and can also explain the baryon to entropy ratio in the universe.  Moreover, density fluctuations in the IBH gas might lead to merger and coalescence of a tiny fraction of the IBHs into more stable PBHs, which can explain the crossover from the radiation to matter dominated eras and plausibly the observed dark matter in the universe.  Thus, all of the required features of the cosmology of the very early universe can be explained by simple models with a small number of parameters.  These models also resolve the conundrum of the initial singularity.  At very early times, the CEP implies that the universe in any causal diamond is a finite dimensional quantum system.  If one views general relativity as a hydrodynamic description of a quantum system obeying the CEP\cite{ted95}, then it is no surprise that this description breaks down when the quantum system is small.  

The largest set of "parameters" in our model is contained in the detailed form of the slow roll metric $H(t) = \frac{\dot{a}}{a} $, which is used to fit the scalar CMB data.
Defining $p$ and $\rho$ by
\begin{equation} H^2 (t)  = (24\pi)^{-1} \rho(t) , \end{equation}
\begin{equation} \dot{\rho} + 3 H \rho = - 3 H p(t) , \end{equation}
we get the standard form of the Friedmann equations for a perfect fluid FRW metric.
The second law of thermodynamics, applied to the CEP, implies\cite{ted95} the null energy condition 
\begin{equation} 0 \leq (p + \rho)\equiv  \dot{\phi}^2 . \end{equation}
As long as $\dot{\phi} \neq 0$ we can write $\rho - p \equiv 2V(\phi)$, defining the "inflationary potential".  So every slow roll metric satisfying the null energy condition/second law is classically equivalent to a single scalar field inflation model.   This shows that single field inflation models, are really models of pure gravity\cite{maldaetal}. The quantum fluctuations in CEP models (see the appendix) do not have the same form as those derived by treating quadratic fluctuations around the slow roll background as quantum fields.  In particular, the scalar power spectrum is larger by a factor of $\epsilon^{-1}$, and this prefactor is the only large violation of $SO(1,4)$ invariance when the inflationary Hubble radius is much larger than the Planck scale.  The primordial tensor fluctuations have two components, an exactly flat spectrum suppressed by $\epsilon^2$ , and gravitons produced in the decay of IBHs.  The latter share the spatial distribution of scalar fluctuations, with a power spectrum smaller by a factor of $g^{-1}$ .  $g$ is the effective number of particle states with mass below $(8\pi m)^{-1} \sim 10^{12}$ GeV.  

   The general principles of CEP cosmology, even with the addition of the "fast scrambling" hypothesis for equilibration times\cite{hpss}, puts only mild restrictions on the form of the slow roll metric.   None of the bounds of the "Swampland" program\cite{vafaetal} apply here.  It is clear that despite the fact that the formulae relating the metric to data are different in detail, the problem of fitting present day CMB observations with CEP models contains exactly the same freedom as fitting a conventional inflationary potential to the data.

Apart from the slow roll metric, the only parameters of these models are the inflationary horizon size, $g$, $g< < g$, $\epsilon_{CP}$, the cosmological constant, and the details of the PBH spectrum.  The latter are in principle calculable in terms of the first three, though the calculations are beyond our present abilities and are subject to the caveats discussed at the end of the last section.  We have taken a value of $g$ consistent with the supersymmetric extension of the standard model and modest extensions thereof, with a SUSY breaking scale less than about $10^{12}$ GeV.  The inflationary horizon size was fit to the CMB spectrum.  It has to be much larger than the Planck scale in order for the statistical considerations of the present paper to make any sense.  Thus, the CMB fluctuations are guaranteed to be small, but their actual size is not determined by any fundamental principle.  

In the context of the CEP, the value of the c.c. is determined by the dimension of the Hilbert space describing the maximal causal diamond in dS space.  It's obviously a model parameter, independent of initial conditions\footnote{Any model of our cosmological history can be embedded in a wide variety of multiverse models, with varying degrees of rigor in their mathematical definition.   Such models can give us a theoretical explanation for the value of the c.c. and/or can lead to instability of dS space on a time scale larger than the dS Hubble radius.  Absent any definite predictions for measurements we could actually do, such models are just intellectual exercises, more in the province of philosophy than of science.  As long as no multiverse model can be distinguished from one in which the c.c. is determined by the dimension of the final Hilbert space, those models can be ignored by the principle of Occam's razor.}.  

Weinberg's bound\cite{w} on the c.c., coming from the requirement that a single galaxy be able to form is (dropping factors of order 1)
\begin{equation} \Lambda < \rho_{eq} ({\frac \delta \rho}{\rho} )^3 =  \rho_{eq} (m \epsilon)^{-3} .\end{equation}
In CEP based models, the values of $m$, and $\epsilon$ appear to be model parameters, while a probability distribution for $\rho_{eq}$ is determined, if we make the assumption that dark matter consists of PBHs.   Weinberg's bound becomes
\begin{equation} \Lambda <  m^{-8} \tilde{g}^2 (m\epsilon)^{-3} (Xr(X))^{4} . \end{equation}  We also recall the constraint coming from insisting that PBHs be stable until the beginning of the matter dominated era
\begin{equation} m^3 X^5 r^2 (X) > g_< . \end{equation}   In the context of CEP based models then, Weinberg's bound becomes a joint constraint on several model parameters.  General principles tell us that $\Lambda^{-1} \ll 1 \ll m$ and that $\epsilon$ is small but not so small as to allow the constrained state defining the dilute gase of IBHs to equilibrate. A quantum probability distribution for $X r(X)$ is, in principle, calculable from the values of the other parameters.   Since construction of a PBH of mass $Xm$ involves the merger of roughly $X$ IBHs, the distribution is likely to favor small values of $X r(X)$.  Weinberg's bound is thus satisfied only in models with very small $\Lambda$. 

\begin{center}
Acknowledgements 
\end{center} 

  We thank Kathryn Zurek for sharing her insights about fluctuations in causal diamonds.  The work of TB is supported in part by the DOE under grant DE-SC0010008.  The work of WF is supported by the NSF under grant PHY-1914679


\begin{thebibliography}{99}
\bibitem{fsb} 
W.~Fischler and L.~Susskind,
``Holography and cosmology,''
[arXiv:hep-th/9806039 [hep-th]];
R.~Bousso,
``A Covariant entropy conjecture,''
JHEP \textbf{07}, 004 (1999)
doi:10.1088/1126-6708/1999/07/004
[arXiv:hep-th/9905177 [hep-th]].
R.~Bousso,
``Holography in general space-times,''
JHEP \textbf{06}, 028 (1999)
doi:10.1088/1126-6708/1999/06/028
[arXiv:hep-th/9906022 [hep-th]];
R.~Bousso,
``The Holographic principle for general backgrounds,''
Class. Quant. Grav. \textbf{17}, 997-1005 (2000)
doi:10.1088/0264-9381/17/5/309
[arXiv:hep-th/9911002 [hep-th]].
\bibitem{bhg}J.~D.~Bekenstein,
``Black holes and entropy,''
Phys. Rev. D \textbf{7}, 2333-2346 (1973)
doi:10.1103/PhysRevD.7.2333
S.~W.~Hawking,
``Black hole explosions,''
Nature \textbf{248}, 30-31 (1974)
doi:10.1038/248030a0
G.~W.~Gibbons and S.~W.~Hawking,
``Cosmological Event Horizons, Thermodynamics, and Particle Creation,''
Phys. Rev. D \textbf{15}, 2738-2751 (1977)
doi:10.1103/PhysRevD.15.2738
\bibitem{tH} G.~'t Hooft,
``Dimensional reduction in quantum gravity,''
Conf. Proc. C \textbf{930308}, 284-296 (1993)
[arXiv:gr-qc/9310026 [gr-qc]].
\bibitem{ted95} T.~Jacobson,
``Thermodynamics of space-time: The Einstein equation of state,''
Phys. Rev. Lett. \textbf{75}, 1260-1263 (1995)
doi:10.1103/PhysRevLett.75.1260
[arXiv:gr-qc/9504004 [gr-qc]].
\bibitem{ted2015} T.~Jacobson,
``Entanglement Equilibrium and the Einstein Equation,''
Phys. Rev. Lett. \textbf{116}, no.20, 201101 (2016)
doi:10.1103/PhysRevLett.116.201101
[arXiv:1505.04753 [gr-qc]].
\bibitem{BDF} T.~Banks, P.~Draper and S.~Farkas,
``Path Integrals for Causal Diamonds and the Covariant Entropy Principle,''
Phys. Rev. D \textbf{103}, no.10, 106022 (2021)
doi:10.1103/PhysRevD.103.106022
[arXiv:2008.03449 [hep-th]].
\bibitem{GH} G.~W.~Gibbons and S.~W.~Hawking,
``Action Integrals and Partition Functions in Quantum Gravity,''
Phys. Rev. D \textbf{15}, 2752-2756 (1977)
doi:10.1103/PhysRevD.15.2752
\bibitem{fs} W.~Fischler and L.~Susskind,
``Holography and cosmology,''
[arXiv:hep-th/9806039 [hep-th]].
\bibitem{holocosmo} T.~Banks, W.~Fischler and L.~Mannelli,
``Microscopic quantum mechanics of the p = rho universe,''
Phys. Rev. D \textbf{71}, 123514 (2005)
doi:10.1103/PhysRevD.71.123514
[arXiv:hep-th/0408076 [hep-th]].
T.~Banks and W.~Fischler,
``Holographic Theories of Inflation and Fluctuations,''
[arXiv:1111.4948 [hep-th]];
T.~Banks and W.~Fischler,
``Holographic Inflation Revised,''
doi:10.1017/9781316535783.013
[arXiv:1501.01686 [hep-th]].
\bibitem{carlipetal} S.~Carlip,
``Black hole entropy from horizon conformal field theory,''
Nucl. Phys. B Proc. Suppl. \textbf{88}, 10-16 (2000)
doi:10.1016/S0920-5632(00)00748-9
[arXiv:gr-qc/9912118 [gr-qc]];
S. Carlip, Phys. Rev. D 51, 632 (1995), arXiv:gr-qc/9409052'S. Carlip, Phys. Rev. Lett. 82, 2828 (1999), arXiv:hep-th/9812013, S. Carlip, Class. Quant. Grav. 15, 3609 (1998), arXiv:hep-th/9806026,S. Carlip, AIP Conf. Proc. 1483, 54 (2012), arXiv:1207.1488 [gr-qc], S. N. Solodukhin, Phys. Lett. B 454, 213 (1999), arXiv:hep-th/9812056, and references therein.

\bibitem{tbkz} T.~Banks and K.~M.~Zurek,
``Conformal Description of Near-Horizon Vacuum States,''
[arXiv:2108.04806 [hep-th]].
\bibitem{hpss} P.~Hayden and J.~Preskill,
``Black holes as mirrors: Quantum information in random subsystems,''
JHEP \textbf{09}, 120 (2007)
doi:10.1088/1126-6708/2007/09/120
[arXiv:0708.4025 [hep-th]]; Y.~Sekino and L.~Susskind,
``Fast Scramblers,''
JHEP \textbf{10}, 065 (2008)
doi:10.1088/1126-6708/2008/10/065
[arXiv:0808.2096 [hep-th]].
\bibitem{mss} J.~Maldacena, S.~H.~Shenker and D.~Stanford,
``A bound on chaos,''
JHEP \textbf{08}, 106 (2016)
doi:10.1007/JHEP08(2016)106
[arXiv:1503.01409 [hep-th]].
\bibitem{gh1} G.~W.~Gibbons and S.~W.~Hawking,
``Cosmological Event Horizons, Thermodynamics, and Particle Creation,''
Phys. Rev. D \textbf{15}, 2738-2751 (1977)
doi:10.1103/PhysRevD.15.2738
\bibitem{tbwfbaryogen} T.~Banks and W.~Fischler,
``CP Violation and Baryogenesis in the Presence of Black Holes,''
[arXiv:1505.00472 [hep-th]].

\bibitem{nz}  Zeldovich, Ya.B., and Novikov, I.D. 1966, Astron. Zh. 43, 758; 1967, Sov.
Astronomy 10, 602
\bibitem{ch} B.~J.~Carr and S.~W.~Hawking,
``Black holes in the early Universe,''
Mon. Not. Roy. Astron. Soc. \textbf{168}, 399-415 (1974)
\bibitem{BD} T.~Banks and P.~Draper,
``Remarks on the Cohen-Kaplan-Nelson bound,''
Phys. Rev. D \textbf{101}, no.12, 126010 (2020)
doi:10.1103/PhysRevD.101.126010
[arXiv:1911.05778 [hep-th]];
\bibitem{maldaetal} A.~R.~Liddle, D.~H.~Lyth, {\it Cosmological Inflation and Large Scale Structure}, Cambridge University Press, 2000; J.~M.~Maldacena,
``Non-Gaussian features of primordial fluctuations in single field inflationary models,''
JHEP \textbf{05}, 013 (2003)
doi:10.1088/1126-6708/2003/05/013
[arXiv:astro-ph/0210603 [astro-ph]], and references therein.
\bibitem{CKN} A.~G.~Cohen, D.~B.~Kaplan and A.~E.~Nelson,
``Effective field theory, black holes, and the cosmological constant,''
Phys. Rev. Lett. \textbf{82}, 4971-4974 (1999)
doi:10.1103/PhysRevLett.82.4971
[arXiv:hep-th/9803132 [hep-th]];
T.~Banks and P.~Draper,
``Remarks on the Cohen-Kaplan-Nelson bound,''
Phys. Rev. D \textbf{101}, no.12, 126010 (2020)
doi:10.1103/PhysRevD.101.126010
[arXiv:1911.05778 [hep-th]];
A.~G.~Cohen and D.~B.~Kaplan,
``Gravitational contributions to the electron $g$-factor,''
[arXiv:2103.04509 [hep-ph]].
\bibitem{constraints} 
B.~Carr, K.~Kohri, Y.~Sendouda and J.~Yokoyama,
``Constraints on Primordial Black Holes,''
[arXiv:2002.12778 [astro-ph.CO]].
B.~Carr and F.~Kuhnel,
``Primordial Black Holes as Dark Matter: Recent Developments,''
Ann. Rev. Nucl. Part. Sci. \textbf{70}, 355-394 (2020)
doi:10.1146/annurev-nucl-050520-125911
[arXiv:2006.02838 [astro-ph.CO]].
\bibitem{ddm}A.~Nygaard, T.~Tram and S.~Hannestad,
``Updated constraints on decaying cold dark matter,''
JCAP \textbf{05}, 017 (2021)
doi:10.1088/1475-7516/2021/05/017
[arXiv:2011.01632 [astro-ph.CO]], and references cited therein.
\bibitem{vafaetal} E.~Palti,
``The Swampland: Introduction and Review,''
Fortsch. Phys. \textbf{67}, no.6, 1900037 (2019)
doi:10.1002/prop.201900037
[arXiv:1903.06239 [hep-th]], and references cited therein.


\bibitem{w} S.~Weinberg,
``Anthropic Bound on the Cosmological Constant,''
Phys. Rev. Lett. \textbf{59}, 2607 (1987)
doi:10.1103/PhysRevLett.59.2607
\end{thebibliography}
\end{document}